\documentclass{llncs}

\let\subparagraph\paragraph
\usepackage{setspace}
\usepackage[T1]{fontenc}
\usepackage{xcolor}
\usepackage{titlesec}
\usepackage{blindtext}
\usepackage[backend=bibtex]{biblatex}
\titlespacing*{\subsection}{0pt}{0.2\baselineskip}{0.2\baselineskip}
\usepackage{graphicx}
\usepackage{epstopdf}
\begin{document}
\title{EEG classification for visual brain decoding with spatio-temporal and transformer based paradigms }
%\thanks{Supported by organization x.}}

\author{Akanksha Sharma\inst{1} \and
Jyoti Nigam \inst{1} \and Abhishek Rathore\inst{1}\and Arnav Bhavsar\inst{1}}
\authorrunning{A. Sharma et al.}
\institute{Indian Institute of Technology-Mandi, Himachal Pradesh, India 
\email{t22110@students.iitmandi.ac.in, jyoti\_nigam@projects.iitmandi.ac.in, abhishek\_rathore@projects.iitmandi.ac.in, arnav@iitmandi.ac.in}\\
}
\maketitle              % typeset the header of the contribution
\begin{abstract}
In this work, we delve into the EEG classification task in the domain of visual brain decoding via two frameworks, involving two different learning paradigms. Considering the spatio-temporal nature of EEG data, one of our frameworks is based on a CNN-BiLSTM model. The other involves a CNN-Transformer architecture which inherently involves the more versatile attention based learning paradigm. In both cases, a special 1D-CNN feature extraction module is used to generate the initial embeddings with 1D convolutions in the time and the EEG channel domains. Considering the EEG signals are noisy, non stationary and the discriminative features are even less clear (than in semantically structured data such as text or image), we also follow a window-based classification followed by majority voting during inference, to yield labels at a signal level. To illustrate how brain patterns correlate with different image classes, we visualize t-SNE plots of the BiLSTM embeddings alongside brain activation maps for the top 10 classes. These visualizations provide insightful revelations into the distinct neural signatures associated with each visual category, showcasing the BiLSTM's capability to capture and represent the discriminative brain activity linked to visual stimuli. We demonstrate the performance of our approach on the updated EEG-Imagenet dataset with positive comparisons with state-of-the-art methods. Our best performing model yields an accuracy of 71\%, a significant improvement over various existing methods.% accuracy for CNN-BiLSTM model and 0.59 accuracy for CNN-transformer model}.
    
%by evaluating two sequential feature extraction techniques—combining Convolutional Neural Networks (CNNs) with Bi-directional Long Short-Term Memory (BiLSTM) networks, and pairing CNNs with Transformer models—applied to the EEG-ImageNet dataset (updated in year 2020). Both approaches surpass current state-of-the-art methods in classification accuracy, demonstrating their efficacy in extracting meaningful patterns from complex EEG signals for visual categorization tasks. Remarkably, the CNN-BiLSTM method outperforms the CNN-Transformer model, challenging the anticipated superiority of Transformers in handling temporal data. This discrepancy raises intriguing questions about the optimal configuration of neural network architectures for EEG signal analysis, suggesting that the nuanced temporal dynamics of EEG data may benefit more from the bidirectional sequential processing offered by BiLSTMs. Our findings contribute to the ongoing discussion on the best practices for EEG data analysis and underscore the importance of empirical testing in uncovering the most effective machine learning models for specific tasks.

\end{abstract}

\section{Introduction}
Electroencephalography (EEG) represents a pivotal tool in neuroscience, enabling the non-invasive measurement of electrical activity in the brain. Its applications span from clinical diagnostics to interfacing with computational systems, exemplifying a bridge between human cognitive functions and machine understanding. In recent years, advancements in signal processing and machine learning have propelled the capabilities of EEG to not just record, but also to decode and classify signals corresponds to the neural activity for various tasked condition, ushering in a new era of brain-computer interfaces (BCIs).

Apart from the more traditional applications such as motor imagery analysis, emotion classification, mental workload analysis etc., EEG-based perceptual brain decoding stands out as a cutting-edge research area. A sub area, involving visual brain decoding amalgamates visual perception with machine learning algorithms to classify and interpret visual stimuli based on brainwave patterns.

Classification of visual stimuli using EEG data are particularly intriguing due to their potential to decode subjective visual experiences without any physical action from the user. This research domain explores how different visual stimuli, such as images or videos, elicit distinct patterns in brain activity that can be classified into predefined categories using machine learning algorithms. Such a task opens up new avenues for applications in areas such as assistive technologies, neuromarketing, and cognitive research.

However, EEG-based visual classification is not without its hurdles. The high non-uniformity of brain signals, combined with the noise and non-stationarity inherent in EEG data, requires sophisticated signal processing and machine learning techniques to achieve accurate classification. Furthermore, the inter-individual variability in EEG signals necessitates adaptive and personalized approaches to model training and testing. Recent deep learning methods for EEG-based visual classification involves various architecture designs providing insights into various aspects covering their suitability for EEG classification.

Convolutional Neural Networks (CNNs) excel at automatically learning spatial hierarchies from visual inputs, making them adept at extracting both spatial and temporal features from EEG data. Their ability to manage high-dimensional data and recognize local dependencies has led to significant success in classifying EEG signals for various visual tasks, including the differentiation of specific visual stimuli and the decoding of visual attention or intention. However, CNNs come with high computational demands and require extensive labeled datasets to mitigate the risk of overfitting.

On the other hand, Recurrent Neural Networks (RNNs) and their more sophisticated variant, particularly Long Short-Term Memory (LSTM) Networks, are capable of capturing long-term dependencies within time-series data. This makes them highly effective for continuous EEG data classification, such as tracking cognitive state or attention shifts over time. Yet, the challenges of training complexity, potential for overfitting, and interpreting the models' internal mechanisms remain notable concerns.

Hybrid models combine features of CNNs, RNNs/LSTMs, and sometimes GCNs to leverage the strengths of each in processing EEG signals \cite{zhang2020bi} and can capture both spatial and temporal features efficiently.% They often achieve superior performance by effectively handling the complexities of EEG data, offering a balance between computational efficiency and classification accuracy \cite{}. 
These models are versatile and can be tailored for a wide range of EEG-based visual classification tasks, from basic stimulus categorization to more complex applications like emotion recognition or neurofeedback.

Transformers are models centered around an attention mechanism, which allows them to consider the entire signal collectively rather than in segments. Theoretically, this attention mechanism equips the model with the capability to recognize and leverage long-term dependencies within the data, regardless of the sequence's length. This feature is particularly advantageous in EEG signal analysis, where understanding the broader context and connections across extensive data sequences is crucial for accurate classification.

The main challenge lies in designing an good architecture that balances the contribution of each component model. Additionally, training hybrid models can be computationally demanding and require careful tuning. In our study, we suggest adopting a convolution neural network as feature extractor at first level followed by sequential/attention based temporal learning following which we employ two different frameworks: Bidirectional Long Short Term Memory (BiLSTM) and Transformer architectures for visual EEG classification.

The domain of visual brain decoding is relatively recent, the existing work in this area primarily employs either CNN architectures or sequential architectures such as Bi-LSTM and transformers. Our method, however, integrates both CNN and sequential architectures (Bi-LSTM and transformer), thereby utilizing both local and temporal features for classification. 
Our work demonstrates that such frameworks, while common to other areas, can benefit EEG brain decoding, considering that CNN filter embeddings help in learning features from noisy EEG data, followed by sequential and attention based architectures, which exploit natural temporal dependencies in EEG signals. 

Thus, the contributions of this work involve:
\begin{itemize}
    \item 1D CNN feature extraction treating the time and the channel dimensions of EEG separately. % Augmentation of data for proper learning
    \item Modular architecture involving a CNN based feature learning followed by two paradigms of sequence-based learning (Bi-LSTM and Transformer) %temporal and spatial information from raw EEG signals as sequential features
    \item Window based classification considering the non-stationarity of EEG signals, followed by a majority voting to achieve signal level labeling. 
    \item Positive comparisons with various approaches, and analysis involving t-SNE and brain mappings to analyze the discrimination of learned embeddings.
\end{itemize}% Subsequently, the extracted features are classified through a combination of independent component analysis (ICA) and support vector machine (SVM), offering a more comprehensive analysis by considering temporal dynamics in both directions.

\section{Related Work}

Our classification task includes EEG signals which are sequence of electrical signals captured over time through electrodes. Traditional methods done on EEG such as spectrogram, continous wavelet transform are used to extract features from raw EEG \cite{thodoroff2016learning}, \cite{tsinalis2016automatic}. Different classifiers such as k-nearest neighbor, multi layer perceptron and Support Vector Machine (SVM) are also used for classification \cite{li2020perils}. After the introduction of neural networks, Convolutional Neural Network (CNN) \cite{li2020perils}, \cite{craik2019deep} and Recurrent Convolutional Neural Network (RCNN) \cite{bashivan2015learning} are also used for classification. People have also used Long Short Term Memory (LSTM) \cite{li2020perils}, \cite{bozal2017personalized} and Recurrent Neural Network (RNN) \cite{craik2019deep} for classification based on EEG.

For our task to classify EEG signals to one of 40 different classes of ImageNet, people have opted various methods to approach the problem. SyncNet \cite{li2017targeting} and EEGNet \cite{lawhern2018eegnet} are two popular models based on CNN architecture. Li et al. \cite{li2017targeting} developed SyncNet, a model that employs structured 1D convolution layers to adeptly extract information from both the time and frequency domains, subsequently classifying the data through the integrated use of 1D convolutional neural networks (CNNs). On another front, Lawhern et al. \cite{lawhern2018eegnet} introduced EEGNet, which leverages 2D CNNs across different dimensions of EEG data. \cite{zhang2020bi} has converted EEG signals to grayscale images then performed the classification task by extracting features from it. Many papers have used Siamese network \cite{palazzo2020decoding}and similar approach \cite{spampinato2017deep}, \cite{shimizu2022improving} to first bring the feature space of EEG and its corresponding image to
same latent space using CNN or RNN and then use classifier for classification of EEG signals.

Since EEG is a sequence, an architecture rich in storing sequential data should be considered. RNNs are prone to vanish gradient for large sequences. Therefore, we have opted transformer and Bidirectional Long Short Term Memory (Bi-LSTM).

There are different existing methods using this approach. \cite{tao2021gated} has changed the residual connection in transformer after attention and feed forward network with GRU or SigTanh calling it as Gated Transformer. \cite{xie2022transformer} has used transformer with CNN in different combination to capture different information such as channel, temporal and fusion of both with different types of positional embeddings. \cite{fares2019eeg} used Bi-LSTM to learn features and then select them using independent component analysis (ICA) for classifying using SVM.

Li et al. \cite{li2017targeting} also critiqued the methodology employed by Spampinato et al. \cite{spampinato2017deep}, pointing out that their reported results were contingent on a block design approach that may not hold up under a rapid-event design process. Specifically, they observed that the division of training and testing sets in Spampinato et al.'s study ensured that every trial in the test sets was derived from a block that was well-represented in the corresponding training set. Li et al. further argued that this methodology inadvertently boosted classification accuracy by capturing the long-term brain activity linked to the blocks of trials, rather than accurately reflecting the brain's response to individual class stimuli. 

Palazzo et al. \cite{palazzo2020decoding} revisited their earlier research to address the criticisms raised by Li et al. \cite{li2017targeting}. This introspection revealed that the original classification performance, which they had reported with an average accuracy of around 83\%, was indeed inflated.

By applying correct filters, particularly targeting the high-frequency gamma-band and more realistic settings, the accuracy of classification showed a marked difference: their method achieved nearly 20\% accuracy. This adjustment not only highlighted the significance of proper data filtering but also vindicated their block design approach for classification studies. 

This discovery implies that the models developed in previous studies cannot be directly compared due to their reliance on unfiltered EEG data, while the filtered dataset was only made available in 2020. In this study, our analysis and comparative evaluations are exclusively focused on methodologies that have utilized the filtered version of the EEG-ImageNet dataset. \cite{mishra2022eeg} has generated results on new dataset for approaches existing on old dataset which shows that EEGNet \cite{lawhern2018eegnet} reported about 30\%, and EEG-Channel Net \cite{palazzo2020decoding} saw a jump to approximately 50\% accuracy.

All the existing approach on our task utilizes either only CNN architectures or only sequential architectures to find only local features or only sequential features from EEG signal. Our approach on contrary, uses both local as well as sequential features extracted by CNN and sequential architecture (Bi-LSTM or transformer) combinedly.

While our CNN-transformer architecture is similar to the Conformer architecture \cite{song2022eeg}, there are key differences:
The Conformer architecture with its potential to first finds local temporal features and then spatial features which are then averaged before giving it to the transformer. This averaging is very well suitable for some well established EEG tasks such as motor imagery and emotion recognition paradigms, where the discrimination is relatively simpler with high accuracy in literature, whereas in our CNN-transformer which uses windows of EEG signal in its approach, averaging will result in loss of information. Therefore, the the first layer of convolution does noise removal and local temporal and spatial features are calculated by further layers for visual stimulus recognition, arguably a harder problem. 

%\textcolor{blue}{Our CNN-transformer architecture is also much simpler than complex architectures like Vision Transformers. While the concept of learning local embeddings followed by a transformer model is similar, Vision Transformers \cite{han2022survey} first split the image into fixed-size patches and then linearly embed these patches. The embedded patches are then processed using a standard transformer encoder, which captures global relationships through multi-head self-attention mechanisms.}

Our architecture is much simpler than complex architectures like Vision Transformers (ViT) \cite{han2022survey}. While the concept of learning local embeddings followed by a transformer model is similar, ViT first split the image into fixed-size patches and then linearly embed these patches which are given to encoder of the Transformer. While in our CNN-Transformer approach, the 1D CNN layers learn local spatio-temporal contexts. On the other hand, ViT architecture flatten the patches to form sequence. While CNN-Transformer finds local temporal features of each channel to form a sequence. The architecture identifies sequential features, where each sequence represents local temporal features, and each local temporal feature contains spatial information across all channels.

\section{Dataset}
In our research, we utilized the updated and filtered EEG dataset released in 2020, a pioneering dataset specifically designed for multi-class visual classification tasks using ImageNet. For ease of reference, this dataset will henceforth be referred to as EEG-ImageNet.

This dataset comprises EEG recordings from six participants who were exposed to a subset of the ImageNet dataset, encompassing 40 different classes with each class containing 50 images. One image per class is shown in Figure \ref{fig:img}. The EEG data was captured using 128 electrodes at a sampling rate of 1000 Hz, with each recording lasting 500 ms. According to the findings of Kaneshiro et al., the initial 500 ms of EEG responses in single trials hold significant information about the categories and attributes of the visual objects under study. Interestingly, they also noted that a mere 80 milliseconds of data from a single electrode could suffice for the classification of EEG signals.

 \begin{figure}[ht]
    \centering
    \includegraphics[width=\linewidth]{croppedImages.png}
    \caption{40 classes present in EEG-ImageNet dataset}
    \label{fig:img}
\end{figure}

From an initial collection of 12,000 recordings, we identified 11,964 valid trials, excluding 36 samples due to their low quality. Additionally, our review revealed the absence of 11 trials for one specific class (mushrooms, identified as class 33) for one participant. Consequently, we excluded all data related to class 33 from both the Image and EEG datasets, which resulted in a refined dataset comprising 39 classes with a total of 11,682 samples.
    
%Several versions of the EEG-ImageNet dataset were prepared, applying different bandpass filters—ranging from 5-95 Hz to 14-70 Hz—to accommodate various experiments. This allowed us to explore the dataset through multiple spectral lenses, capturing the nuances of theta, alpha, beta, and gamma brainwave information during visual stimulation. For analytical consistency and to facilitate comparison across samples, data normalization was performed using a z-score transformation for each channel, ensuring zero-centered values with a standard deviation of one.

\section{Methodology }
\begin{figure}[ht]
\centering
\vspace{-10mm}
\includegraphics[width=1.2\linewidth, height=0.9\linewidth]{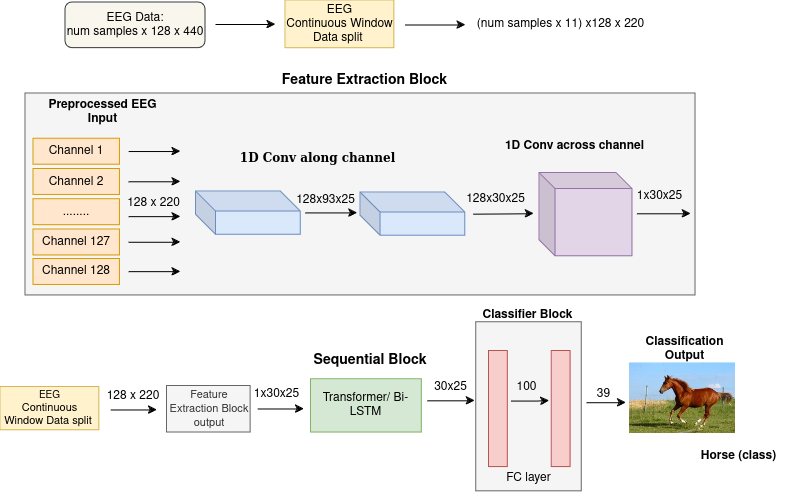}
\caption{The architecture diagram shows the following steps: An EEG sample is divided into $11$ segments of $220$ samples each during pre-processing. This is followed by a Feature Extraction block which uses a CNN to capture spatial and temporal information. then a Sequential block, which can be a Bi-LSTM or Transformer, learns sequential relations. Finally, a Classifier block with dense layers to classify the embeddings}
\vspace{-5mm}
\label{fig:OverallArchitecture}
\end{figure}

In this work, we employ two different frameworks with an initial CNN head to compute EEG embedding. These are then fed to a sequential block (Bi-LSTM in one of the frameworks) and Transformer block in another. Our architecture of the model is shown in Figure \ref{fig:OverallArchitecture}.

In the model, the feature extractor is specifically designed considering the nature of EEG signal. The Bi-LSTM/Transformer block learns relationship among features fed to it. Finally, a classification block operates on the output of Bi-LSTM/Transformer block to compute class labels. 

The architecture offers flexibility in modifying either component according to the specific requirements of the EEG classification task or the characteristics of the EEG data, making it a versatile choice for researchers.
%\blindtext[2]

The raw EEG data is not directly provided to feature extractor block of the model, but instead passes through a preprocessing step. The preprocessing step also serves to augment the data for the model. We discuss all the above in detail in the subsections below.

\subsection{Pre-processing}
The dataset used contains frequencies from 5-95Hz. A notch filter at 50Hz is also used around the power line. For analytical consistency, data normalization was performed using a z-score transformation for each channel, ensuring zero-centered values with a standard deviation of one.

Further, the pre-processing also involves windowing where EEG data of each image is divided into smaller time windows. Initially, the overall EEG data per image, for each of the EEG channel is of length 440. Windows of 220 time samples of EEG data are extracted from this 440 length of EEG data with 90\% overlapping in consecutive windows. This creates 11 windows of 220 length from single 440 length EEG data, all having same label as that of original signal. The windowing also leads to data augmentation, while considering the non-stationarity in EEG signal.

\subsection{Feature Extraction Block}
This block contains CNN to separately extract spatial and temporal features across EEG channels and time axis. It performs 1-D convolution for three layers. 1-D convolution not only contains less parameters as compared to 2-D convolution but also helps in extracting temporal features from each channel and spatial features across channels, explicitly. This provides more control over generating embeddings in temporal and channel information without intermixing them. This is well explained in \cite{mishra2021eeg} and \cite{mishra2020analyzing}. The first two layers perform convolution along time (activation function = Relu, kernel size = (1,35) and stride = (1,2)) while last layer do convolution over channel (activation function = sigmoid, kernel size = (128,1) and stride = (1,1)). The convolution layers over time find features across different time stamps while convolution layer over channel finds features across all channels for each extracted feature of time stamp. Each convolution layer has 25 filters. The final output of feature extractor is reshaped to (30,25) and fed to sequential block.

\subsection{Bi-LSTM and Transformer Blocks}
The extracted CNN features are provided as input to the next stage are of size (30,25), which should be treated as 30 sequences each having feature vector of 25 units. In order to learn the relationships among the CNN embeddings for such a sequence, we used two different approaches: using Bi-LSTM and using Transformer.

\textbf{Bi-LSTM Module :}
%The EEG multi-channel temporal signal which has its temporal dynamics, ignoring temporal nature might lead to loss of valuable information, to capture this temporal information RNN’s types of sequence model can be used.
LSTM  which is variant of RNN, has a capability to capture sequential dependencies in the input data. Bi-LSTM which contains two LSTM one is to track information in forward direction another one to track information in backward direction, processing of data in forward direction and backward direction capture more meaningful correlations in EEG signals. Bi-LSTM can  be used as encoder to translate EEG signals from EEG space to low dimensional vector space which act as representation of EEG signal. This representation can further be used for classification stage.

EEG signal can be fed to LSsTM in two ways: either channel\cite{spampinato2017deep} wise or all EEG channel vector per time step. In common LSTM architecture whole EEG signal per time step is fed to LSTM. In Stacked LSTM, multiple layer of LSTM are used. Output of first layer is given to next layer. The final output of last layer at the last time step is the encoding of EEG signal\cite{mishra2022eeg}, which can further be used for classification. In the Bi-LSTM block, we have stacked Bi-LSTM to capture information in both forward and backward direction.

\begin{figure}[ht]
\centering
\vspace{-10mm}
\includegraphics[scale = 0.55]{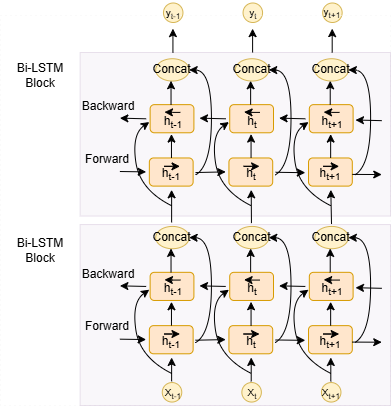}
\caption{Architecture of stacked Bi-LSTM in sequential block which is fed with extracted features in both forward and reverse manner in order to learn sequence in both forward and backward direction. Present feature is represented as $x_t$ while previous feature is represented as $x_{t-1}$.}
\vspace{-5mm}
\label{fig:Bi-LSTM Architecture}
\end{figure}
In the model, LSTM is used as recurrent unit for Bi-LSTM with 22 storing units. The sequential block of Bi-LSTM used is shown in Figure \ref{fig:Bi-LSTM Architecture}. Two Bi-LSTMs are stacked together to form the sequential block.

\textbf{Transformer based EEG model :}
In order to learn long non-local relationships in EEG embeddings, we used another version of our framework, where the CNN embeddings are followed by a transformer block.

The Transformer model, consisting of an encoder and decoder architecture for sequence to sequence mapping. The encoder processes the input data utilizing self-attention mechanisms to weigh the non-local importance among different parts of data. This processed information is then passed to the decoder, which also employs self-attention and additionally uses what is known as cross-attention to focus on relevant parts of the input sequence when generating new output sequence. This allows Transformers to efficiently handle sequences of data, capturing complex relationships within data compared to recurrent layers.

\begin{figure}[ht]
\centering
\vspace{-10mm}
\includegraphics[width=0.6\linewidth]{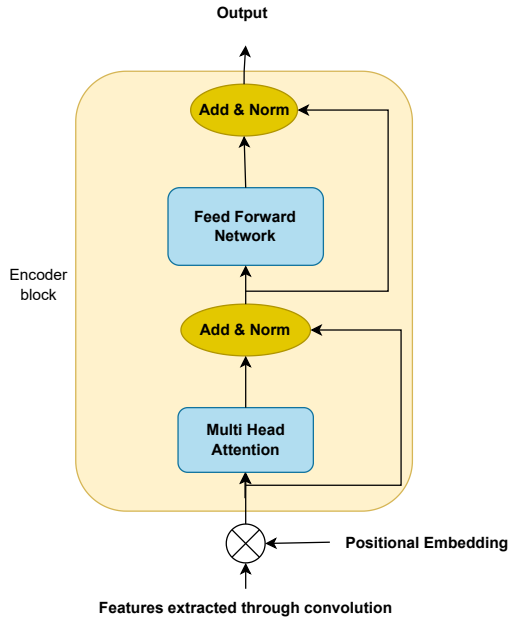}
\caption{Architecture of Transformer for sequential block. It consist of only encoder part with single encoder layer. The extracted features from feature extractor is fed to it to give embedding which are used for classification.}
\vspace{-5mm}
\label{fig:TransformerArchitecture}
\end{figure}

The Architecture mentioned in \cite{vaswani2017attention} consist of encoder and decoder for sequence to sequence mapping. As our take only involves EEG sequence classification, we therefore use only the encoder part of the transformer. The encoder of the transformer for our approach finds non-local relations within the sequential feature embeddings provided by the CNN using self-attention. The proposed architecture as shown in Figure \ref{fig:TransformerArchitecture} has 1 encoder layer with 8 heads for multi-head attention and 2 dense layers in feed forward network.
%\textcolor{red}{Conversely, while Transformers excel at processing long-range dependencies and parallelizing computation, their performance may be influenced by the specific characteristics of EEG data, such as its high dimensionality, noise levels, and the subtlety of temporal patterns related to visual stimuli.}

%\blindtext[3]

\subsection{Classifier Block}
The embeddings from Bi-LSTM/Transformer blocks are finally used as the input to the classifier block which consists of sequential relationship among features.A dense layer of fully connected neural network is used for classification. It consists of 100 neurons with sigmoid as activation function. After this, another layer with 39 output neurons corresponding to the 39 classes is used with softmax as its activation function.

\subsection{Maximum Voting}
As mentioned in Section 4.1, the proposed architecture uses windows for training, validation and testing. This leads to classification at the level of windows. This does not classify whole EEG signal corresponding to single image, but instead assigns labels to different windows of the sequence. In order to label whole EEG signal, we employ majority voting across the windows. The class label which is predicted by majority of windows from a single EEG sequence is assigned to the whole sequence.

\section{Discussion on some salient features of the modules}
The selection of deep learning modules for EEG classification is based on intuitive considering both spatial and temporal nature of EEG. While CNNs offer robust feature extraction for learning local spatio-temporal relationship analysis, LSTMs excel in capturing temporal dynamics. For long sequences, Transformer works on attention paradigm, which involves extracting non local relationships. Below we discuss such salient features of such modules. %   which overcome this challenge and find temporal relations even for long sequences. Transformer performs parallel computations which makes it fast.

\subsection{CNN}
% Combining Convolutional Neural Networks (CNNs) with Bi-directional Long Short-Term Memory (Bi-LSTM) networks presents a powerful architecture for extracting temporal features from EEG data for visual classification tasks. Below are the salient aspects of this architecture:

 %\begin{itemize}
  %   \item{Feature Extraction }
\begin{enumerate}
     \item Spatial and Temporal Feature Extraction: CNNs are renowned for their ability to extract high-level, abstract features from spatial data. In the context of EEG data, CNN layers can efficiently identify spatial patterns and relationships for both across different EEG channels and time stamps, which are crucial for understanding the underlying brain activity related to visual stimuli.
 %\item{Improved Accuracy and Contextual Understanding} 
  %\item{Robustness and Generalization }
     \item Noise Reduction: The initial CNN layers act as a form of automatic feature engineering, which can help in reducing noise from the raw EEG signals before they are processed for temporal feature extraction. This preprocessing step can improve the robustness of the model against variations in the EEG signal quality.
% \item Generalization Capability: The combination of CNN and Bi-LSTM networks can generalize well to new, unseen EEG data, thanks to the model's ability to learn both spatial and temporal features. This makes the architecture versatile across different subjects and experimental conditions. 
 
%\textbf{ Computational Efficiency }
 
 %A. Efficient Training: Despite the complexity, models combining CNN and Bi-LSTM layers can be trained relatively efficiently due to the ability of CNNs to reduce the dimensionality of the input data before it reaches the Bi-LSTM layers, thus making the training process faster and more computationally efficient. 
 
% \item{ Flexibility and .
 
% \item Scalability: This architecture can be easily scaled up or down by adjusting the number and size of the CNN and Bi-LSTM layers to accommodate different sizes of datasets and computational constraints.Overall, the combination of CNNs and Bi-LSTMs for EEG-based visual classification tasks leverages the strengths of both spatial and temporal feature extraction techniques, offering a comprehensive approach to understanding and classifying brain activity related to visual stimuli.
 \end{enumerate}
%\end{itemize}

\subsection{BiLSTM}
\begin{enumerate}
    \item Temporal Dynamics: Bi-LSTM layers complement CNNs by capturing temporal dependencies and dynamics within the EEG signals. Unlike traditional RNNs, Bi-LSTMs process data in both forward and backward directions, offering a richer understanding of temporal sequences by incorporating both past and future context.
     \item Contextual Understanding: The bidirectional nature of Bi-LSTM networks enables a broader contextual understanding, potentially enhancing the accuracy of classifying visual tasks from EEG data, especially in scenarios where accurate predictions depend on the sequence of brain activity. Moreover, Bi-LSTMs are specifically designed to manage long-term dependencies, allowing them to retain information over extended periods which is crucial for EEG data, where the significance of signal features can extend across diverse time frames. This capability makes Bi-LSTMs particularly effective for EEG-based tasks that require detailed temporal analysis. 
 \end{enumerate}

\subsection{ Transformer}
%Utilizing a combination of Convolutional Neural Networks (CNNs) followed by Transformer models to extract temporal features from EEG data for visual classification tasks represents a cutting-edge approach. This architecture leverages the strengths of both models to handle complex spatial and temporal data, offering several salient features:
\begin{enumerate}
    %\item Robust Spatial Feature Identification: CNNs excel at extracting hierarchical spatial features from EEG data, effectively identifying patterns and relationships across both  the EEG channels and time stamps that are indicative of specific visual stimuli. This ability makes CNNs an ideal first step in processing EEG data for visual classification.

%\textbf{Enhanced Feature Extraction}
\item Non-local Temporal Dynamics: Transformers introduce a sophisticated mechanism for capturing temporal relationships in data. Unlike traditional sequence processing models, Transformers use self-attention mechanisms to weigh the importance of different parts of the input data, allowing for a more nuanced understanding of temporal sequences within the EEG signals.
\item Global Context Awareness: The self-attention mechanism in Transformers enables the model to consider the entire sequence of data at once, thereby capturing global dependencies. This global perspective is particularly beneficial for EEG data, where the significance of a signal might be dependent on the entire sequence of brain activity.
\item Effectiveness in long sequence data: EEG data streams are typically long sequences. Transformers, with their self-attention mechanism, are adept at handling such long sequences without losing performance, making them suitable for EEG-based tasks.
\end{enumerate}
%}
%\textbf{Superior Contextual Interpretation}

%B. Efficient Training and Inference: Despite their complexity, Transformers can be efficiently trained, especially when combined with CNNs for initial feature extraction. This efficiency makes it feasible to deploy these models for real-time classification tasks.In summary, leveraging a CNN followed by a Transformer model for EEG-based visual classification tasks harnesses the strengths of both architectures, offering a sophisticated and efficient approach to understanding the intricate patterns of brain activity associated with visual stimuli. This combination not only enhances the model's performance in classifying visual tasks but also offers flexibility, scalability, and computational efficiency, making it a promising architecture for advanced EEG analysis.

\section{Experiments and Results}
\subsection{Experimental Setup}

For our experiment, the split of 80\% training, 10\% validation and 10\% test data is used as done in \cite{tao2021gated}. Each split contains data of each class for each subject. EEG data of all subjects for single image of a particular class is also taken into single split (either training or validation or testing) for uniformity. While splitting data it was observed that 33rd class has irregular data among different subjects. To maintain regularity among subject and class data, the 33rd class is removed, this resulting in a total of 39 classes.

%The model was not able to learn for split data as it got stuck to very shallow local minima while optimizing the problem. The performance of a model designed using deep learning depends on the point where it converges (local or global minima). This convergence depends on various hyper parameters such as initial values of kernels, kernel initializer, learning rate, batch size etc. for a single architecture. Its performance can also be improved by changing parameters such as number of dense layers for classification, number of sequential units, etc. 
Our experimentation includes both parameter and hyper parameter tuning including initial values of filters, filter initializer learning rate, batch size, number of neurons in dense layer, number of dense layers, number of encoder layers for Transformer, number of storing units of Bi-LSTM. The two models are run for 400 and 760 iterations with difference between loss of successive iterations less than 0.0001 as convergence criteria.
\singlespacing

\subsection{Results} 
The result of the architectures and comparison with other existing work is shown in Table \ref{tab:comp}. These are either from their respective papers or from \cite{mishra2022eeg}, which reports a comparison of various approaches.\par\par
\begin{table}
    \centering
    \caption{Comparision on results of Existing Work}
    \begin{tabular}{|c|c|} \hline 
         \textbf{Approach}& \textbf{Accuracy (\%)}\\ \hline 
         Stacked Bi-LSTM \cite{fares2019eeg} & 22\\ \hline 
          EEGCVPR40\cite{spampinato2017deep}& 26\\ \hline 
         SyncNet \cite{li2017targeting}& 27\\ \hline 
         %\cite{palazzo2020decoding}& 31\\ \hline 
 EEGNet \cite{lawhern2018eegnet}&32\\ \hline 
 EEG-ChannelNet \cite{palazzo2020decoding}&36\\ \hline 
 GRU Gated Transformer \cite{tao2021gated}&46\\ \hline 
 SigTanh Gated Transformer \cite{tao2021gated}&49\\ \hline 
 EfficientNet + SVM \cite{mishra2022eeg}&64\\ \hline
 \textbf{Proposed Bi-LSTM approach}&\textbf{71}\\\hline
 \textbf{Proposed Transformer approach}&\textbf{59}\\\hline
    \end{tabular}
    \label{tab:comp}
\end{table}
%Out of our two approaches used, Bi-LSTM has learned training data well than transformer and exhibit more generalisation for test data.\par

%It was inferred that the features given to sequential block contains information that can be well interpreted by sequential block for single encoder layer and fails to capture information for more number of encoder layers in transformer. This is because of providing features to sequential block instead of giving data directly to it as feature embeddings provide  more important information from data to sequential block. This avoids not only large number of parameters but also complexity of architecture. Even embeddings given to sequential block made Bi-LSTM to perform well due to richness in information in the input of Bi-LSTM. \par
%90-10 split model is over seen data which cause restriction for its generalization. Training for 80-10-10 split makes our model more general. The graphs in Figure \ref{fig:transformer} and Figure \ref{fig:Bi-LSTM} ensures about learning of features and relationships present in data by both models without any bias to any splits of data. The generalisation of our models give better results over existing methods to approach the classification problem. Tabel \ref{tab:comp} shows the comparision among existing works.

%\par
We observe that our CNN-Bi-LSTM based method outperforms all existing approaches and the CNN-Transformer based method also outperform the most and better than all existing Transformer based approaches. Both our approaches have taken the leverage of both local as well as sequential features to perform the task while most existing approaches are limited to either only local features or sequential features only. The Bi-LSTM approach even performs better over approach in \cite{zhang2020bi} which involves converting EEG signals to gray-scale image. \par
\begin{table}
    \centering
    \caption{Performance metrics for the proposed Bi-LSTM model}
    \begin{tabular}{|c|c|c|c|} \hline 
         \textbf{Accuracy}& \textbf{Recall}& \textbf{Precision}& \textbf{F1-Score}\\ \hline 
         71\%& 71\%& 72\%& 71\%\\ \hline 
      %   SigTanh Gated Transformer& 49\\ \hline 
       %  Our Transformer approach& 59\\ \hline
    \end{tabular}
    \label{tab:otherMetrics}
\end{table}
We note that, presently the CNN-Transformer methods (ours as well as other methods) have low performance compared to the CNN-Bi-LSTM approach. We believe that a reason for this % relatively low performance of Transformer based design 
could be due to smaller window length, since Transformers usually expects long duration data to better learn non-local dependencies. Moreover, the architecture and training dynamics of Transformers might necessitate adjustments for optimal performance on EEG datasets. Factors such as the choice of hyper parameters, the design of the attention mechanism, and the amount of training data available could significantly impact the effectiveness of Transformers as this is an unconventional application domain. Hence, the unique challenges posed by EEG data require bespoke modifications to the standard Transformer model to fully leverage its capabilities.

Having said that, from Table \ref{tab:comp}, we also note that the Transformer based framework in this work outperforms the other existing Transformer based approaches which uses a gated network (using either GRU Gate or SigTanh Gate) in place of the residual network of the encoder of Transformer. Thus, this work highlights the potential of using Transformer based architectures in this domain.

 \begin{figure}[ht]
    \centering
    \vspace{-5mm}
    \includegraphics[width=0.8\linewidth]{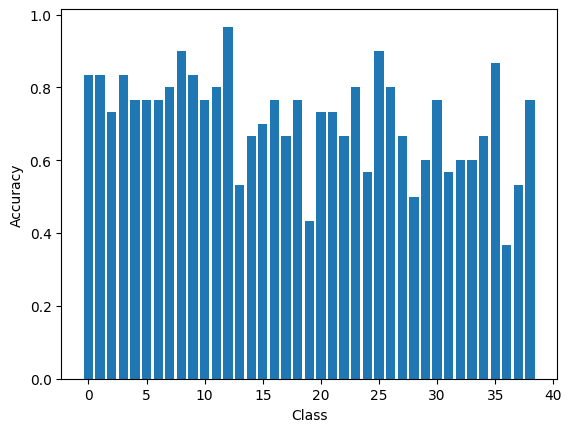}
    \caption{Bar graph for accuracy of individual classes}
    \vspace{-5mm}
    \label{fig:accClss}
\end{figure}

Table \ref{tab:otherMetrics} and Figure \ref{fig:accClss} also provide further results for our CNN-Bi-LSTM model which shows that the performance is consistently distributed for most classes, and the method achieve high scores for precision, recall and F1-score, which highlights the recognition of individual classes.
% based on transformer while Table \ref{tab:comp} shows that our Bi-LSTM approach outperforms all other existing approaches for EEG-based classification.

%In our investigation into EEG-based visual classification, we meticulously evaluated two sequential feature extraction methodologies: the first combining Convolutional Neural Networks (CNNs) with Bi-directional Long Short-Term Memory (BiLSTM) networks, and the second pairing CNNs with Transformer models. Both approaches yielded classification accuracies surpassing current state-of-the-art methods,  underscoring the effectiveness of leveraging advanced neural network architectures for analyzing EEG data. Intriguingly, our findings revealed that the CNN-BiLSTM framework outperformed the CNN-Transformer configuration in terms of accuracy, a result that warrants further exploration given the conventional wisdom regarding Transformers' superior handling of temporal data. 
\singlespacing
\subsection{Analysing embeddings and brain mappings}

Given the superior performance of the CNN-BiLSTM model, we opted to use this model to conduct an analysis focusing on the relationship between embeddings and brain mappings, as an initial step towards interpreting the model's performance.

Typically, in many works involving neuroimage/signal analysis, such topographical maps are used to highlight differences between conditions. In this case the conditions correspond to the classes. While the t-SNE visualizations demonstrate the difference in the class embeddings from the neural networks, the topographical maps serve a complementary purpose of suggesting, that there are also differences in the spatial distribution of brain activity across the classes. When associated with t-SNE plots, this leads to an insightful hypothesis that such original differences in the signal patterns can get encoded in the network embeddings. This additional layer of analysis would contribute to the explainability of such methods by linking brain activation patterns to the classified visual stimuli.
 
\begin{figure}[ht]
\centering
\vspace{-5mm}
\includegraphics[scale = 0.5]{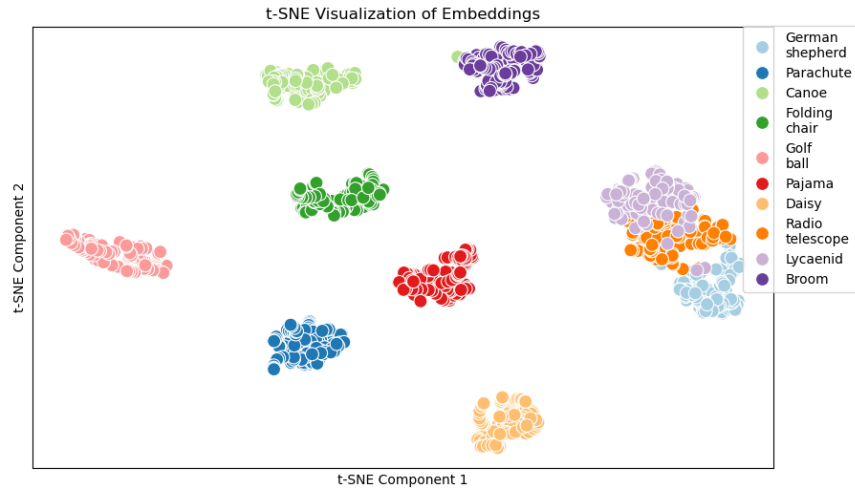}
\caption{Embedding Visualization of Top 10 classes based on Accuracy}
\vspace{-5mm}
\label{fig:bilstm_tsne}
\end{figure}

% \begin{figure}[h]
% \centering
% \includegraphics[scale = 0.5]{ICPR_2024_LaTeX_Templates/images/bilstm_embeddings_plot_2.png}
% \caption{Embedding Visualization of Top 10 classes based on Accuracy}
% \label{fig:bilstm_tsne}
% \end{figure}

We visualized the embeddings extracted from the last BiLSTM layer. Our analysis specifically targeted the top 10 performing classes for Table \ref{tab:top_10acc}. Utilizing t-SNE plots of embeddings corresponding to 100 random window samples for each class (Fig \ref{fig:bilstm_tsne}), we observed distinct patterns in the latent space. These patterns formed well-defined clusters that corresponded to the various classes, indicating a significant separation and organization within the embedding space. This clustering strongly supports the high classification accuracy observed within these groups.

\begin{table}
    \centering
    \caption{Top 10 Classes based on Accuracy}
    \begin{tabular}{|c|c|c|c|c|c|}
    \hline
     \textbf{Class} & German Shepherd & Parachute & Canoe & Folding Chair & Golf ball 
     \\
     \hline
     \textbf{Accuracy} & 98.87 & 95.1 & 95.07 & 94.44 & 94.37 \\
     \hline
    \textbf{Class} & Pajama & Daisy & Egyptian Cat & Lycaenid & Broom \\
     \hline
      \textbf{Accuracy} & 93.24 & 93.12 & 92.85 & 92.62 & 92.39 \\
    \hline
    \end{tabular}
    %\vspace{6mm}
    \label{tab:top_10acc}
\end{table}
% In the analysis of the CNN-BiLSTM model, the BiLSTM layer's embeddings were visualized, revealing distinctive patterns in the latent space. The top 10 classes, as detailed in Table \ref{tab:top_10acc}, showcased the highest levels of accuracy. Further, when these embeddings were mapped onto a two-dimensional plane using t-SNE, as depicted in fig \ref{fig:bilstm_tsne}, well-defined clusters emerged that corresponded to the different classes. Such clustering suggests a pronounced separation and organization within the embedding space, corroborating the classification accuracy observed for these groups.
 % table  shows  the top 10 classes  with the highest accuracies are depicted , showcasing the similarity in latent space embeddings learned by the BiLSTM layer. This representation is visually presented in the accompanying figure \ref{fig:bilstm_tsne}
The superior performance of the CNN-BiLSTM model could be attributed to several factors. BiLSTM's strength in capturing both forward and backward temporal dependencies may offer a more nuanced understanding of EEG signal sequences, which are inherently complex and non-linear. This bidirectional processing potentially provides a more comprehensive context for each point in the sequence, crucial for the good quality classification of EEG data.

% \textcolor{red}{Conversely, while Transformers excel at processing long-range dependencies and parallelizing computation, their performance may be influenced by the specific characteristics of EEG data, such as its high dimensionality, noise levels, and the subtlety of temporal patterns related to visual stimuli.}

% \textcolor{red}{Moreover, the architecture and training dynamics of Transformers might necessitate adjustments for optimal performance on EEG datasets. Factors such as the choice of hyperparameters, the design of the attention mechanism since CNN is applied beforehand, and the amount of training data available could significantly impact the effectiveness of Transformers in this context. It is also possible that the unique challenges posed by EEG data require bespoke modifications to the standard Transformer model to fully leverage its capabilities for temporal feature extraction.}

\begin{figure}[ht]
\centering
\vspace{-5mm}
\includegraphics[scale = 0.3]{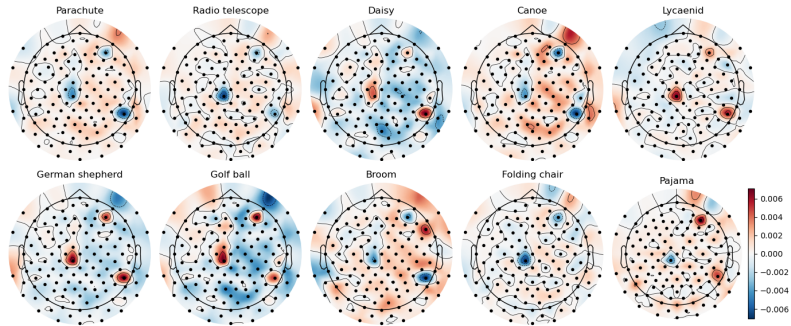}
\caption{Topographic Maps of EEG Signal Amplitudes for 10 Classes}
\vspace{-5mm}
\label{fig: topomap}
\end{figure}

% \vspace{-15pt}
As the counterpart of the deep learning embeddings,  the topographic brain maps in Figure \ref{fig: topomap} provides an overview of the active EEG signal mean amplitudes across ten distinct classes, offering insights into the spatial distribution of neural activity spanning on a wide frequency spectrum from 5 to 95 Hz. This analysis encompasses prominent EEG frequency bands such as alpha (8-13 Hz), beta (13-30 Hz), and gamma (30-70 Hz), which are associated with various cognitive processes and neurological conditions.

\par The maps also show distinct pattern among different classes. Thus, it seems plausible that such these discriminative topographic plots among classes are being learnt well resulting in distinct cluster in embedding space and then resulting in good performance.

%These results underscore the importance of considering the specificities of EEG data when selecting and optimizing machine learning models for visual classification tasks. The CNN-BiLSTM approach, with its emphasis on sequential temporal dynamics, presents a compelling option for EEG analysis, though further research is needed to unravel the nuanced interplay between model architecture, data characteristics, and task requirements. Additionally, these findings invite a deeper examination of how Transformers can be tailored or enhanced to better capture the temporal complexities inherent in EEG signals, ensuring that their theoretical advantages are fully realized in practical applications.

\section{Conclusion}
In this study, we proposed two approaches for EEG-based visual classification by integrating Convolutional Neural Networks (CNNs) separately with Bi-directional Long Short-Term Memory (Bi-LSTM) networks and Transformer models, both of which significantly outperformed existing state-of-the-art methods. The CNN-BiLSTM methodology excelled in capturing both the spatial and temporal dynamics of EEG data, harnessing Bi-LSTM's prowess in modeling long-term dependencies over time. Alternatively, the CNN-Transformer approach leveraged the attention mechanism of Transformers to interpret complex temporal relationships. A breif analysis was also provided towards to end considering t-SNE visualization and brain mappings, which offers some insight into the discrimination of neural patterns and the learnt embeddings. These results not only demonstrate the potential of leveraging advanced deep learning architectures for more reliable EEG-based visual classification but also set a new benchmark in brain-computer interface research. Moving forward, we hope to build on this advances towards achieving more generalization across diverse settings, and better interpretability connecting neural and deep learning aspects.

\subsubsection{Acknowledgements} The work has been partly supported by Science and Engineering Research Board, Government of India. 
%\bibliographystyle{splncs04}
%\bibliography{references}
\printbibliography
\end{document}